\documentclass[12pt,preprint]{aastex}
\newcommand{\HI}{H\,{\sc i}}
\newcommand{\Gff}{{G5.4$-$1.2}}
\newcommand{\Gft}{{G5.27$-$0.9}}
\newcommand{\psr}{{PSR B1757$-$24}}
\shorttitle{PSR~B1757$-$24 and G5.4$-$1.2}
\begin{document}
\title{Decapitating the Duck: The (non)association of PSR~B1757$-$24
  and supernova remnant G5.4$-$1.2}
\author{S. E. Thorsett}
\affil{Department of Astronomy and Astrophysics, University of
  California, Santa Cruz 95064}
\email{thorsett@ucolick.org}
\author{W. F. Brisken and W. M. Goss}
\affil{National Radio Astronomy Observatory, P.O. Box 0,
 Socorro, New Mexico 87801}
\begin{abstract}
  We have made the first direct interferometric proper motion
  measurements of the radio pulsar B1757$-$24, which sits at the tip
  of the ``beak'' of the putative ``Duck'' supernova remnant. The
  peculiar morphology of this radio complex has been used to argue
  alternately that the pulsar's space motion was either surprisingly
  high or surprisingly low.  In fact, we show that the pulsar's motion
  is so small that it and its associated nonthermal nebula G5.27$-$0.9
  (the ``head'') are almost certainly unrelated to the much larger
  G5.4$-$1.2 (the ``wings'').
\end{abstract}
\section{Introduction}
The radio source colloquially known as the ``Duck'' (Fig.~1) consists
of the large (35\arcmin\ diameter) fan-shaped nebula \Gff\ with edge
brightening along the western edge (the ``wing''); the small
(1\arcmin\ diameter) nebula \Gft\ a few arcminutes to the west (the
``head''); and a narrow (10\arcsec\ wide) extension reaching a further
30\arcsec\ west (the ``beak'').  At the very western tip 
is the 125~ms radio pulsar \psr, with a
characteristic spin-down age $\tau=P/2\dot P=15.5$~kyr.

Even before the localization of the pulsar or the measurement of its
small apparent age, the Duck drew attention for its peculiar
morphology \citep{hb85,ckk+87}. The pulsar's youth and location near
but outside the remnant only intensified the interest.  The large
displacement of the pulsar from its putative birthplace has twice led
to conclusions sufficiently remarkable to be widely reported in the
press: first the apparent youth of the pulsar led to the
conclusion it was traveling at $\sim2000$~km/s \citep{fk91,mkj+91},
then a surprisingly small limit on the motion of the pulsar bow shock
led to the conclusion that the pulsar must be far older than its
characteristic age \citep[henceforth GF00]{gf00}.

The fundamental assumption on which these and scores of other
publications depend is that the Duck is a coherent structure, not
a chance line-of-sight juxtaposition of unrelated sources. This
hypothesis allows several predictions: First, the distances to \Gff,
\Gft, and \psr\ should all agree; second, the pulsar and remnant ages
should agree; finally, the pulsar should be moving in a ballistic
trajectory away from the supernova blast center.  Testing these
predictions has been hard. The distance estimates from \HI\ absorption
measurements and pulsar dispersion are not inconsistent, but are poorly
constrained; the age estimates, initially in excellent agreement, have
proven inconsistent with the proper motion limits; and the proper
motion inferred from the bow shock is in the wrong direction.

Our goal was a direct measurement of the pulsar motion, to help
elucidate the nature of the Duck.

\section{The proper motion measurement}
High precision interferometric astrometry of \psr\ is difficult
because the pulsar is buried within a bright, extended radio
nebula. GF00 used limits on the motion of the extended source as a
proxy for motion of the pulsar, but faced significant uncertainties
about the time variability of the extended emission and the
homogeneity of the interstellar medium near the pulsar. 

VLBI observations of B1757$-$24 have been attempted by some of us, but
have been unsuccessful because of the low pulsar flux density and the
large distance ($\sim3^\circ$) to the nearest VLBI calibrator,
J1751$-$253, which is strongly scattered (to $\sim40$~mas at 1.4~GHz).
Instead, we made observations with the Very Large Array, in A array at
20~cm wavelength. Isolation of the pulsar from the extended nebula was
achieved by gating the VLA correlator synchronously with the pulsar
period, to accept data only when the pulsar was ``on.''

Observations were made on 1998 Jun 3, 2002 Feb 20, and 2002 May 2. For
the latter two epochs, the VLBA antenna at Pie Town, New Mexico, was
added using the newly available fiber link, improving resolution in
the east-west direction.  Astrometry was done by measuring the pulsar
position relative to seven compact reference sources (fluxes
2.5--90~mJy) in the same field.  Reference source positions were
determined to between 3 and 50~mas. These data acquisition and
analysis techniques were identical to that used for our large pulsar
proper motion study, and have been discussed in much greater depth in
\citet{mbf+01}.

The gated image is shown in Fig.~1. Our estimate (95\% confidence) of
the proper motion is: $(\mu_\alpha, \mu_\delta)=(-2.1\pm14.0,
-14\pm25)$ mas/yr.  The J2000.0 position at epoch 2002.334, measured
relative to the geodetic calibrator 1751$-$253, is RA 18:01:00.023(3),
Dec $-$24:51:27.53(5), displaced more than $5\sigma$ south of the
position given in GF00 but in excellent agreement with the position of
the X-ray point source detected by \citet{kggl01}.

Our proper motion measurement is consistent with the motion inferred
by GF00 from studies of the ``bow shock'': $(\mu_\alpha,
\mu_\delta)=(5.8\pm12.2, -1.3\pm10.1)$ mas/yr, where again we quote
95\% confidence limits.\footnote{Note that the sign of $\mu_\alpha$
  was incorrectly stated in GF00---they in fact measured an eastward
  motion at the $1\sigma$ level, as their analysis makes clear
  (Gaensler and Frail, private communication).}  
We combine these independent results for a best (95\%) estimate of
$(\mu_\alpha, \mu_\delta)=(2.4\pm9.2, -3.1\pm9.4)$ mas/yr.  Our upper
limit on the {\it westward} motion of the pulsar is therefore
$\sim6.8$~mas/yr.

\section{The case for association}

\subsection{Distance}
It is often stated that \Gff\ and \Gft\ have consistent \HI\ distance
measurements.  In fact, the data \citep*{fkw94} are unconstraining. The
\HI\ absorption profiles of the two sources are essentially identical,
with absorption detectable out to $+27$~km/s.  With a standard
galactic rotation model, this gives a a distance limit $d>4.3$~kpc for
both objects. It must be emphasized that only a lower limit is
available: the essentially identical absorption measurement for
1757$-$248, an extragalactic source in the field, implies there is no
absorbing material at greater distances.  \HI\ measurements alone
cannot imply the \Gff\ and \Gft\ are at the same distance any more
than they imply \Gff\ and 1757$-$248 are at the same distance.

A pulsar distance estimate comes from the measured radio dispersion
together with a galactic electron density model.  The standard
\citet{tc93} electron model yields $d\sim 4.6$~kpc, with an
uncertainty of perhaps 40\% \citep{bri01}.  The new electron model of
\citet*[preferred secant model]{gbc01} gives a range of 5.8--13~kpc,
with a most likely value of 9.1~kpc.


\subsection{Age}
The pulsar is 21~arcminutes from the geometric center of \Gff, but the
apparent line of motion of the pulsar does not point back to this
center (\S\ref{sec:dir}). Attempting to model the required asymmetric
expansion of the remnant, \citet{fkw94} found possible birth locations
16.1--20.6\arcmin\ east of the pulsar position.  With our westward
proper motion limit, this requires that the pulsar is at least
140-180~kyr old (with the younger ages requiring an extreme asymmetry
in the remnant expansion, \S\ref{sec:dir}).

Another estimate of the pulsar age is given by the spin-down timescale
$$\tau=\frac{-\nu}{\left(n-1\right)\dot\nu}
\left[1-\left(\frac{\nu}{\nu_0}\right)^{n-1}\right],$$
where $\nu_0$
is the initial spin frequency and the spin-down torque has been
assumed to be proportional to a power of the spin frequency,
$\dot\nu\propto\nu^n$. The braking index $n$ would be three for
magnetic dipole dominated spin-down; the handful of available observed
values range from two to three, except for the Vela pulsar with
$n=1.4\pm0.2$ \citep[see review by][]{ls98}.

The ``characteristic age,'' assuming $n=3$ and $\nu_0\gg\nu$, is
$P/2\dot P=15,470$~yr \citep{mkj+91}, far smaller than the kinematic
age derived by assuming an association with \Gff.

The timing and proper motion age estimates cannot be reconciled for
any spin-down model with braking index $n>1$. For any initial spin
frequency\footnote{Pulsar initial spin frequencies are known in only a
  few cases, including the Crab pulsar, $\nu_0\approx50$~Hz
  \citep{mt77}, and the pulsar in G11.2$-$0.3, $\nu_0\approx16$~Hz
  \citep{krv+01}.  There has been considerable recent investigation
  of the possibility that r-mode instabilities and their associated
  gravitational radiation may rapidly drive nascent neutron stars to
  frequencies $\nu\lesssim100$~Hz \citep[reviewed by][]{ak01}.}
$\nu_0<100$~Hz and braking index $n>1$, the true age of the pulsar is
$\lesssim75$~kyr ($\lesssim130$~kyr if $\nu_0<500$~Hz). For a braking
index consistent with the smallest measurement for any pulsar
($n=1.2$, consistent with the Vela lower bound), and an almost
certainly physically unreasonable initial spin frequency near the
neutron star limit, $\nu_0\sim2$~kHz, the age is only 100~kyr.



Instead, it is necessary to posit a model with increasing torque or
decreasing moment of inertia.  For example, a magnetic field that is
exponentially growing on a $\sim30$~kyr timescale could explain the
discrepancy \citep*{bah83}.

Unfortunately, the remnant age is not sufficiently constrained to
provide an independent estimate. We note that the close coincidence
between the 14~kyr estimated age of \Gff\ \citep{ckk+87} and the
15.5~kyr characteristic age of the pulsar was originally considered a
strong argument in favor of the association, but the remnant expansion
timescale can be adjusted widely with adjustments of the assumed
medium density or the distance. Still, if the pulsar and remnant were
born at the same time, it remains a remarkable coincidence that they
independently had their ages initially mis-estimated by the same
factor of ten.

\subsection{Proper motion direction\label{sec:dir}}

The narrow finger of radio and X-ray emission behind the pulsar bow
shock, presumed collimated by the pulsar motion, runs nearly exactly
east-west, and the inferred line of proper motion misses the center of
\Gff\ by 5\arcmin. Although the brightness distribution
across the remnant is highly asymmetric, the remnant is well described
by a circle, with no signs of a breakout in any direction
\citep{fkw94}. If the association is correct, then the blast site must
be 5\arcmin\ north of the current center of symmetry, and the
supernova must be expanding highly asymmetrically (twice as fast to
the south as to the north) without breaking the remnant's symmetry.
The blast site can also be shifted a few arcminutes to the west,
towards the current pulsar location, if an even more asymmetric
expansion is allowed.  \citet{fkw94} present a model of expansion into
a gas layer with a strong exponential density distribution (scale
length about a third to a fourth of the remnant diameter).  However,
the astrophysical difficulty of arranging a factor 20--40 density
gradient across the remnant without local inhomogeneities that disturb
the circular symmetry seems daunting. We consider that the proper
motion direction remains a serious problem for the association
hypothesis.

\subsection{Angular proximity}

The angular proximity of \Gff\ and \Gft\ appears striking, but they
lie in a crowded region of sky near the Galactic center. The
significance of the association is difficult to estimate {\it a
  posteriori}, but we note that the most recent supernova catalog
\citep{gre01} contains 17 SNR within $5^\circ$ of \psr, filling about
2.6~deg$^2$, or just over 3\% of the surface area. About 5\% of the
sky in this region is as close to a remnant as \Gft\ is to \Gff.  Also
in this $5^\circ$ circle are 26 known radio pulsars, so it would be
surprising if there were not at least one or two spurious angular
associations with remnants. (See also \citet{gj95a}.) We conclude that
the positional evidence for association between \psr\ and \Gff\ is
suggestive, but not strong. On the other hand, \psr\ is centered on
the head of the ``bow shock'' extension from \Gft, to within
arcseconds, and we consider this association secure.

\subsection{Edge brightening of \Gff}
Perhaps the most compelling evidence for an association between the
pulsar and \Gff\ has been the relative brightening of the west edge of
the remnant, interpreted as the result of relativistic particles
streaming back along open field lines from the pulsar to re-energize
the shell \citep{ckk+87,sfs89}.  Further evidence comes from
observations of the radio spectral index, which varies from a flat
value ($-0.1$) characteristic of emission from a pulsar-driven,
Crab-like remnant near the pulsar to a steeper value ($-0.4$) more
characteristic of an SNR shell $\sim50^\circ$ away in azimuthal angle,
with systematic uncertainties introduced by the differing
sensitivities of the high and low frequency observations to extended
structure \citep{fkw94}.

The evidence seems less significant when \Gff\ is compared with
similar remnants in similar Galactic environs.  Many supernova remnants
show a significant brightness gradient, increasing towards the
Galactic plane \citep{cl79,dgg+96}, which is to the north-west of
\Gff.  In fact, \Gff\ is one of the ``significant majority'' of
large-diameter, shell-like remnants which shows this effect
\citep{cas77}.  Although the whole western side of the shell is
brighter than the eastern side, the northwest---moving directly toward
the plane---is substantially brighter than the southwest.  Spectral
index variations of a few tenths are also seen in other shell remnants
\citep{ar93,dvgh00}, with flatter spectra typically (but not always)
in brighter regions, just as seen for \Gff.

\section{Summary}
We believe that the evidence linking \psr\ to the remnant \Gff\ is
weak, consisting mainly of the relatively close projected proximity of
the sources together with the unusual edge-brightening of the remnant.
We have suggested that the former is not particularly
surprising in this crowded part of the Galaxy, and the latter may be
due to the brightening of the edge moving into the denser medium along
the Galactic plane.

Standing against this evidence are two difficulties.  First, we must
understand the serious discrepancy between the small timing age of the
pulsar and its long kinematic timescale.  \psr\ now appears to be a
prototypical ``young'' pulsar, with bright radio and X-ray emission
from its associated pulsar wind nebula. If we assume that the pulsar
is really $\sim150$~kyr old, as required if the association is
correct, then we must conclude that it spent the first 90\% or more of
its life looking much ``older'' and less energetic before being
``rejuvenated'' by some substantial increase in its spin-down rate
$\sim10^5$ years ago. While there seems no way to rule out the
possibility that \psr\ was born with a weak magnetic field that has
grown only recently, this appears to us to require unnatural
fine-tuning.  We note that the magnetic field of \psr\ inferred from
its spin-down rate, $4\times10^{12}$~G, is not unusual for a young
pulsar: it is within 10\% of the magnetic field inferred for the
950~yr old Crab pulsar.

Second, we must understand how the remnant could have retained its
circular symmetry over $1.5\times10^5$~yrs while expanding two to
three times faster to the south/south-east and decelerating to a
velocity of a few hundred km/s or less.  Again, this appears to
require unnatural fine-tuning of the parameters of the medium into
which the remnant is expanding, with an extreme gradient in one
direction and extreme uniformity in the transverse direction.

In summary, the unified model originally designed to explain the
angular proximity of \Gff, \Gft, and \psr\ and the morphology of \Gff\ 
has failed both straightforward predictive tests initially proposed,
and can be saved only by positing unusual evolution of {\it both} the
pulsar {\it and} the remnant.  If instead we reject the hypothesis of
a physical association, we find a more straightforward path. We
conclude that \psr\ was most likely born about 15~kyr ago in or near
the Crab-like remnant \Gft---which it is escaping at $\sim 5$~mas/yr
(120~km/s at 5~kpc) and which it continues to power.  No new or
unusual pulsar or remnant physics is required, nor any unusual
properties for the local interstellar medium.  \Gff\ is an unrelated
foreground or background remnant, whose not implausible angular
association with \Gft\ in the dense Galactic center region of the sky
is made more visually spectacular by its edge brightening toward the
Galactic plane.  The Duck is no more a coherent body than is its
northern cousin, Cygnus the Swan.

\acknowledgments The authors thank Dale Frail and Bryan Gaensler for
discussions of their work, and an anonymous referee for helpful
suggestions.  SET is support by NSF grant AST-0098343.
This research has made use of the SIMBAD database, operated at CDS,
Strasbourg, France, and of NASA's ADS Bibliographic Services. The VLA
is operated by the National Radio Astronomy Observatory, a facility of
the NSF operated under cooperative agreement by Associated
Universities, Inc.


\clearpage

\begin{figure}
  \plotone{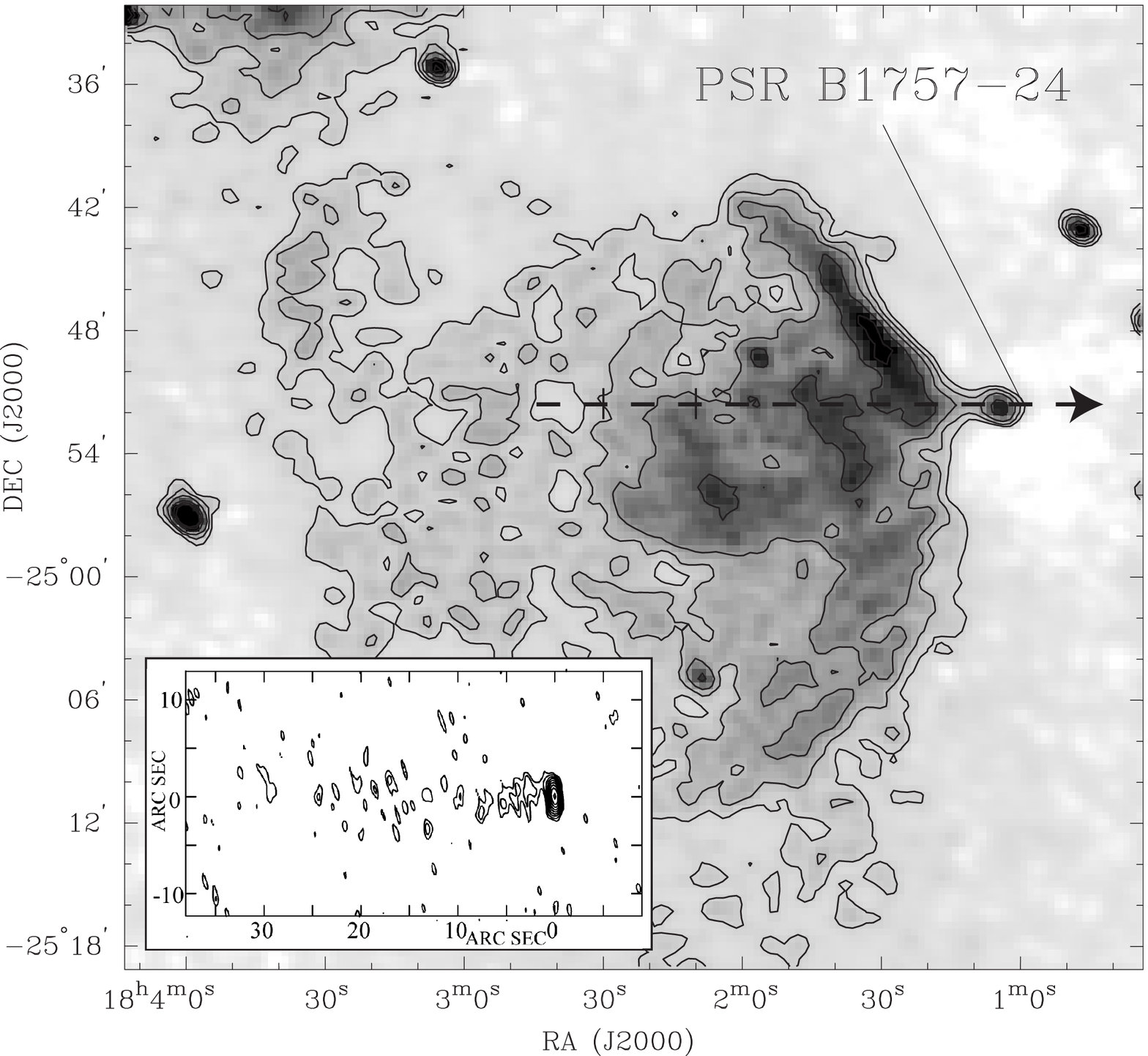} 
  
  \caption{The Duck.  Gray scale image (reprinted with permission
    from GF00) shows the entire complex at 90~cm.  The large, circular
    remnant is \Gff, the small remnant to the west is \Gft, and \psr\
    (not visible) lies as indicated.  The resolution of the image is
    $60\arcsec\times45\arcsec$.  Other information about the
    observations is in GF00.  The dashed line and arrow indicate the
    direction of motion of the pulsar as inferred from the bow shock
    structure in \Gft. Vertical marks indicate the range of possible
    pulsar birth locations that \citet{fkw94} found consistent with
    the remnant symmetry.
    The inset shows the 21~cm image of \psr, using data from the
    VLA and the Pie Town VLBA antenna with the correlator
    gated synchronously with the pulsar period.  The resolution is
    $0.8\arcsec\times2.1\arcsec$, the pulsar image is consistent with a
    point source. The rms noise is about 0.1~mJy/beam.}
\end{figure}
\end{document}